# Eigenvalue based Spectrum Sensing Algorithms for Cognitive Radio *

Yonghong Zeng, *Senior Member, IEEE,* and Ying-Chang Liang, *Senior Member, IEEE*
Institute for Infocomm Research, A*STAR, Singapore.

November 23, 2009

## Abstract

Spectrum sensing is a fundamental component is a cognitive radio. In this paper, we propose new sensing methods based on the eigenvalues of the covariance matrix of signals received at the secondary users. In particular, two sensing algorithms are suggested, one is based on the ratio of the maximum eigenvalue to minimum eigenvalue; the other is based on the ratio of the average eigenvalue to minimum eigenvalue. Using some latest random matrix theories (RMT), we quantify the distributions of these ratios and derive the probabilities of false alarm and probabilities of detection for the proposed algorithms. We also find the thresholds of the methods for a given probability of false alarm. The proposed methods overcome the noise uncertainty problem, and can even perform better than the ideal energy detection when the signals to be detected are highly correlated. The methods can be used for various signal detection applications without requiring the knowledge of signal, channel and noise power. Simulations based on randomly generated signals, wireless microphone signals and captured ATSC DTV signals are presented to verify the effectiveness of the proposed methods.

## 1  Introduction

A "Cognitive Radio" senses the spectral environment over a wide range of frequency bands and exploits the temporally unoccupied bands for opportunistic wireless transmissions [1, 2, 3]. Since a cognitive radio operates as a secondary user which does not have primary rights to any pre-assigned frequency bands, it is necessary for it to dynamically detect the presence of primary users. In December 2003, the FCC issued a Notice of Proposed Rule Making that identifies cognitive radio as the candidate for implementing negotiated/opportunistic spectrum sharing [4]. In response to this, in 2004, the IEEE formed the 802.22 Working Group to develop a standard for wireless regional area networks (WRAN) based on cognitive radio technology [5]. WRAN systems will operate on unused VHF/UHF bands that are originally allocated for TV broadcasting services and other services such as wireless microphone, which are called primary users. In order to avoid interfering with the primary services, a WRAN system is required to periodically detect if there are active primary users around that region.

As discussed above, spectrum sensing is a fundamental component in a cognitive radio. There are however several factors which make the sensing problem difficult to solve. First, the signal-to-noise ratio (SNR) of the primary users received at the secondary receivers may be very low. For example, in WRAN, the target detection SNR level at worst case is $-20$dB. Secondly, fading and time dispersion of the wireless channel may complicate the sensing problem. In particular, fading will cause the received signal power fluctuating dramatically, while unknown time dispersed channel will cause coherent detection unreliable [6, 7, 8]. Thirdly, the noise/interference level changes with time which results in noise uncertainty [9, 6, 10]. There are two types of noise uncertainty: receiver device noise uncertainty and environment noise uncertainty. The sources of receiver device noise uncertainty include [6, 10]: (a) non-linearity of components; and (b) thermal noise in components, which is non-uniform and time-varying. The environment noise uncertainty may be caused by transmissions of other users, including near-by unintentional transmissions and far-away intentional transmissions. Because of the noise uncertainty, in practice, it is very difficult to obtain the accurate noise power.

There have been several sensing algorithms including the energy detection [11, 9, 12, 6, 10], the matched filtering [12, 6, 8, 7] and cyclostationary detection [13, 14, 15, 16], each having different operational requirements, advantages and disadvantages. For example, cyclostationary detection requires the knowledge of cyclic frequencies of the primary users, and matched filtering needs to know the waveforms and channels of the primary users. On the other hand, energy detection does not need any information of the signal to be detected and is robust to unknown dispersive channel. However, energy detection relies on the knowledge of accurate noise power, and inaccurate estimation of the noise power leads to SNR wall and high probability of false

---

alarm [9, 10, 17]. Thus energy detection is vulnerable to the noise uncertainty [9, 6, 10, 7]. Finally, while energy detection is optimal for detecting independent and identically distributed (iid) signal [12], it is not optimal for detecting correlated signal, which is the case for most practical applications.

To overcome the shortcomings of energy detection, in this paper, we propose new methods based on the eigenvalues of the covariance matrix of the received signal. It is shown that the ratio of the maximum or average eigenvalue to the minimum eigenvalue can be used to detect the the presence of the signal. Based on some latest random matrix theories (RMT) [18, 19, 20, 21], we quantify the distributions of these ratios and find the detection thresholds for the proposed detection algorithms. The probability of false alarm and probability of detection are also derived by using the RMT. The proposed methods overcome the noise uncertainty problem and can even perform better than energy detection when the signals to be detected are highly correlated. The methods can be used for various signal detection applications without knowledge of the signal, the channel and noise power. Furthermore, different from matched filtering, the proposed methods do not require accurate synchronization. Simulations based on randomly generated signals, wireless microphone signals and captured digital television (DTV) signals are carried out to verify the effectiveness of the proposed methods.

The rest of the paper is organized as follows. In Section II, the system model and some background information are provided. The sensing algorithms are presented in Section III. Section IV gives theoretical analysis and finds thresholds for the algorithms based on the RMT. Simulation results based on randomly generated signals, wireless microphone signals and captured DTV signals are given in Section V. Also some open questions are presented in this section. Conclusions are drawn in Section VI. A pre-whitening technique is given in Appendix A for processing narrowband noise. Finally, a proof is given in Appendix B for the equivalence of average eigenvalue and signal power.

Some notations used in the paper are listed as follows: superscripts $T$ and $\dagger$ stand for transpose and Hermitian (transpose-conjugate), respectively. $\mathbf{I}_q$ is the identity matrix of order $q$.

## 2 System Model and Background

Let $x_c(t) = s_c(t) + \eta_c(t)$ be the continuous-time received signal, where $s_c(t)$ is the possible primary user's signal and $\eta_c(t)$ is the noise. $\eta_c(t)$ is assumed to be a stationary process satisfying $\mathrm{E}(\eta_c(t)) = 0$, $\mathrm{E}(\eta_c^2(t)) = \sigma_\eta^2$ and $\mathrm{E}(\eta_c(t)\eta_c(t+\tau)) = 0$ for any $\tau \neq 0$. Assume that we are interested in the frequency band with central frequency $f_c$ and bandwidth $W$. We sample the received signal at a sampling rate $f_s$, where $f_s \geq W$. Let $T_s = 1/f_s$ be the sampling period. For notation simplicity, we define $x(n) \triangleq x_c(nT_s)$, $\bar{s}(n) \triangleq s_c(nT_s)$ and $\eta(n) \triangleq \eta_c(nT_s)$. There are two hypothesizes: $H_0$, signal does not exist; and $H_1$, signal exists. The received signal samples under the two hypothesizes are given respectively as follows [6, 8, 7]:

$$H_0: \quad x(n) = \eta(n), \tag{1}$$
$$H_1: \quad x(n) = \bar{s}(n) + \eta(n), \tag{2}$$

where $\bar{s}(n)$ is the received signal samples including the effects of path loss, multipath fading and time dispersion, and $\eta(n)$ is the received white noise assumed to be iid, and with mean zero and variance $\sigma_\eta^2$. Note that $\bar{s}(n)$ can be the superposition of signals from multiple primary users. It is assumed that noise and signal are uncorrelated. The spectrum sensing or signal detection problem is to determine if the signal exists or not, based on the received samples $x(n)$.

**Note:** In above, we have assumed that the noise samples are white. In practice, if the received samples are the filtered outputs, the corresponding noise samples may be correlated. However, the correlation among the noise samples is only related to the receiving filter. Thus the noise correlation matrix can be found based on the receiving filter, and pre-whitening techniques can then be used to whiten the noise samples. The details of a pre-whitening method are given in Appendix A.

Now we consider two special cases of the signal model.

(i) Digital modulated and over-sampled signal. Let $s(n)$ be the modulated digital source signal and denote the symbol duration as $T_0$. The discrete signal is filtered and transmitted through the communication channel [22, 23, 24]. The resultant signal (excluding receive noise) is given as [22, 23, 24]

$$s_c(t) = \sum_{k=-\infty}^{\infty} s(k)h(t - kT_0), \tag{3}$$

where $h(t)$ encompasses the effects of the transmission filter, channel response, and receiver filter. Assume that $h(t)$ has finite support within $[0, T_u]$. Assume that the received signal is over-sampled by a factor $M$, that is, the sampling period is $T_s = T_0/M$. Define

$$\begin{aligned} x_i(n) &= x((nM + i - 1)T_s), \\ h_i(n) &= h((nM + i - 1)T_s), \\ \eta_i(n) &= \eta_c((nM + i - 1)T_s), \end{aligned} \tag{4}$$

$$n = 0, 1, \cdots; \ i = 1, 2, \cdots, M.$$

We have

$$x_i(n) = \sum_{k=0}^{N} h_i(k)s(n - k) + \eta_i(n), \tag{5}$$

where $N = \lceil T_u/T_0 \rceil$. This is a typical single input multiple output (SIMO) system in communications. If there are multiple source signals, the received signal turns out to be

$$x_i(n) = \sum_{j=1}^{P} \sum_{k=0}^{N_{ij}} h_{ij}(k) s_j(n-k) + \eta_i(n), \qquad (6)$$

where $P$ is the number of source signals, $h_{ij}(k)$ is the channel response from source signal $j$, and $N_{ij}$ is the order of channel $h_{ij}(k)$. This is a typical multiple input multiple output (MIMO) system in communications.

(ii) Multiple-receiver model. The model (6) is also applicable to multiple-receiver case where $x_i(n)$ becomes the received signal at receiver $i$. The difference between the over-sampled model and multiple-receiver model lies in the channel property. For over-sampled model, the channels $h_{ij}(k)$'s (for different $i$) are induced by the same channel $h_j(t)$. Hence, they are usually correlated. However, for multiple receiver model, the channel $h_{ij}(k)$ (for different $i$) can be independent or correlated, depending on the antenna separation. Conceptually, the over-sampled and multiple-receiver models can be treated as the same.

Model (2) can be treated as a special case of model (6) with $M = P = 1$ and $N_{ij} = 0$, and $s(n)$ replaced by $\bar{s}(n)$. For simplicity, in the following, we only consider model (6). **Note that the methods are directly applicable to model (2) with $M = P = 1$ (later the simulation for wireless microphone is based on this model)**.

Energy detection is a basic sensing method [11, 9, 12, 6]. Let $T(N_s)$ be the average power of the received signals, that is,

$$T(N_s) = \frac{1}{MN_s} \sum_{i=1}^{M} \sum_{n=0}^{N_s-1} |x_i(n)|^2, \qquad (7)$$

where $N_s$ is the number of samples. The energy detection simply compares $T(N_s)$ with the noise power to decide the signal existence. Accurate knowledge on the noise power is therefore the key to the success of the method. Unfortunately, in practice, the noise uncertainty always presents. Due to the noise uncertainty [9, 6, 10], the estimated noise power may be different from the actual noise power. Let the estimated noise power be $\hat{\sigma}_\eta^2 = \alpha \sigma_\eta^2$. We define the noise uncertainty factor (in dB) as

$$B = \max\{10 \log_{10} \alpha\}. \qquad (8)$$

It is assumed that $\alpha$ (in dB) is evenly distributed in an interval $[-B, B]$ [6, 17]. In practice, the noise uncertainty factor of receiving device is normally 1 to 2 dB [6]. The environment noise uncertainty can be much higher due to the existence of interference [6]. When there is noise uncertainty, the energy detection is not an effective method [9, 6, 10, 17] due to the existence of SNR wall and/or high probability of false alarm.

## 3 Eigenvalue based Detections

Let $N_j \stackrel{\text{def}}{=} \max_i(N_{ij})$. Zero-padding $h_{ij}(k)$ if necessary, and defining

$$\mathbf{x}(n) \stackrel{\text{def}}{=} [x_1(n), x_2(n), \cdots, x_M(n)]^T, \qquad (9)$$
$$\mathbf{h}_j(n) \stackrel{\text{def}}{=} [h_{1j}(n), h_{2j}(n), \cdots, h_{Mj}(n)]^T, \qquad (10)$$
$$\boldsymbol{\eta}(n) \stackrel{\text{def}}{=} [\eta_1(n), \eta_2(n), \cdots, \eta_M(n)]^T, \qquad (11)$$

we can express (6) into a vector form as

$$\mathbf{x}(n) = \sum_{j=1}^{P} \sum_{k=0}^{N_j} \mathbf{h}_j(k) s_j(n-k) + \boldsymbol{\eta}(n), \ n = 0, 1, \cdots. \qquad (12)$$

Considering $L$ (called "smoothing factor") consecutive outputs and defining

$$\hat{\mathbf{x}}(n) \stackrel{\text{def}}{=} [\mathbf{x}^T(n), \mathbf{x}^T(n-1), \cdots, \mathbf{x}^T(n-L+1)]^T,$$
$$\hat{\boldsymbol{\eta}}(n) \stackrel{\text{def}}{=} [\boldsymbol{\eta}^T(n), \boldsymbol{\eta}^T(n-1), \cdots, \boldsymbol{\eta}^T(n-L+1)]^T,$$
$$\hat{\mathbf{s}}(n) \stackrel{\text{def}}{=} [s_1(n), s_1(n-1), \cdots, s_1(n-N_1-L+1), \cdots,$$
$$s_P(n), s_P(n-1), \cdots, s_P(n-N_P-L+1)]^T, \qquad (13)$$

we get

$$\hat{\mathbf{x}}(n) = \mathbb{H}\hat{\mathbf{s}}(n) + \hat{\boldsymbol{\eta}}(n), \qquad (14)$$

where $\mathbb{H}$ is a $ML \times (N + PL)$ ($N \stackrel{\text{def}}{=} \sum_{j=1}^{P} N_j$) matrix defined as

$$\mathbb{H} \stackrel{\text{def}}{=} [\mathbb{H}_1, \mathbb{H}_2, \cdots, \mathbb{H}_P], \qquad (15)$$

$$\mathbb{H}_j \stackrel{\text{def}}{=} \begin{bmatrix} \mathbf{h}_j(0) & \cdots & \cdots & \mathbf{h}_j(N_j) & \cdots & 0 \\ & \ddots & & & \ddots & \\ 0 & \cdots & \mathbf{h}_j(0) & \cdots & \cdots & \mathbf{h}_j(N_j) \end{bmatrix}. \qquad (16)$$

Note that the dimension of $\mathbb{H}_j$ is $ML \times (N_j + L)$.

Define the statistical covariance matrices of the signals and noise as

$$\mathbf{R}_x = \mathrm{E}(\hat{\mathbf{x}}(n)\hat{\mathbf{x}}^\dagger(n)), \qquad (17)$$
$$\mathbf{R}_s = \mathrm{E}(\hat{\mathbf{s}}(n)\hat{\mathbf{s}}^\dagger(n)), \qquad (18)$$
$$\mathbf{R}_\eta = \mathrm{E}(\hat{\boldsymbol{\eta}}(n)\hat{\boldsymbol{\eta}}^\dagger(n)). \qquad (19)$$

We can verify that

$$\mathbf{R}_x = \mathbb{H}\mathbf{R}_s\mathbb{H}^\dagger + \sigma_\eta^2 \mathbf{I}_{ML}, \qquad (20)$$

where $\sigma_\eta^2$ is the variance of the noise, and $\mathbf{I}_{ML}$ is the identity matrix of order $ML$.

### 3.1 The algorithms

In practice, we only have finite number of samples. Hence, we can only obtain the sample covariance matrix other than the statistic covariance matrix. Based on the sample covariance matrix, we propose two detection methods as follows.

**Algorithm 1** *Maximum-minimum eigenvalue (MME) detection*

*Step 1.* Compute the sample covariance matrix of the received signal

$$\mathbf{R}_x(N_s) \stackrel{\text{def}}{=} \frac{1}{N_s} \sum_{n=L-1}^{L-2+N_s} \hat{\mathbf{x}}(n)\hat{\mathbf{x}}^\dagger(n), \quad (21)$$

where $N_s$ is the number of collected samples.

*Step 2.* Obtain the maximum and minimum eigenvalue of the matrix $\mathbf{R}_x(N_s)$, that is, $\lambda_{max}$ and $\lambda_{min}$.

*Step 3. Decision:* if $\lambda_{max}/\lambda_{min} > \gamma_1$, signal exists ("yes" decision); otherwise, signal does not exist ("no" decision), where $\gamma_1 > 1$ is a threshold, and will be given in the next section.

**Algorithm 2** *Energy with minimum eigenvalue (EME) detection*

*Step 1.* The same as that in Algorithm 1.

*Step 2.* Compute the average power of the received signal $T(N_s)$ (defined in (7)), and the minimum eigenvalue $\lambda_{min}$ of the matrix $\mathbf{R}_x(N_s)$.

*Step 3. Decision:* if $T(N_s)/\lambda_{min} > \gamma_2$, signal exists ("yes" decision); otherwise, signal does not exist ("no" decision), where $\gamma_2 > 1$ is a threshold, and will be given in the next section.

The difference between conventional energy detection and EME is as follows: energy detection compares the signal energy to the noise power, which needs to be estimated in advance, while EME compares the signal energy to the minimum eigenvalue of the sample covariance matrix, which is computed from the received signal only.

**Remark:** Similar to energy detection, both MME and EME only use the received signal samples for detections, and no information on the transmitted signal and channel is needed. Such methods can be called *blind detection methods*. The major advantage of the proposed methods over energy detection is as follows: energy detection needs the noise power for decision while the proposed methods do not need.

### 3.2 Theoretical analysis

Let the eigenvalues of $\mathbf{R}_x$ and $\mathbb{H}\mathbf{R}_s\mathbb{H}^\dagger$ be $\lambda_1 \geq \lambda_2 \geq \cdots \geq \lambda_{ML}$ and $\rho_1 \geq \rho_2 \geq \cdots \geq \rho_{ML}$, respectively. Obviously, $\lambda_n = \rho_n + \sigma_\eta^2$. When there is no signal, that is, $\hat{\mathbf{s}}(n) = 0$ (then $\mathbf{R}_s = 0$), we have $\lambda_1 = \lambda_2 = \cdots = \lambda_{ML} = \sigma_\eta^2$. Hence, $\lambda_1/\lambda_{ML} = 1$. When there is a signal, if $\rho_1 > \rho_{ML}$, we have $\lambda_1/\lambda_{ML} > 1$. Hence, we can detect if signal exists by checking the ratio $\lambda_1/\lambda_{ML}$. This is the mathematical ground for the MME. Obviously, $\rho_1 = \rho_{ML}$ if and only if $\mathbb{H}\mathbf{R}_s\mathbb{H}^\dagger = \lambda\mathbf{I}_{ML}$, where $\lambda$ is a positive number. From the definition of the matrix $\mathbb{H}$ and $\mathbf{R}_s$, it is highly probable that $\mathbb{H}\mathbf{R}_s\mathbb{H}^\dagger \neq \lambda\mathbf{I}_{ML}$. In fact, the worst case is $\mathbf{R}_s = \sigma_s^2\mathbf{I}$, that is, the source signal samples are iid. At this case, $\mathbb{H}\mathbf{R}_s\mathbb{H}^\dagger = \sigma_s^2\mathbb{H}\mathbb{H}^\dagger$. Obviously, $\sigma_s^2\mathbb{H}\mathbb{H}^\dagger = \lambda\mathbf{I}_{ML}$ if and only if all the rows of $\mathbb{H}$ have the same power and they are co-orthogonal. This is only possible when $N_j = 0$, $j = 1, \cdots, P$ and $M = 1$, that is, the source signal samples are iid, all the channels are flat-fading and there is only one receiver.

If the smoothing factor $L$ is sufficiently large, $L > N/(M - P)$, the matrix $\mathbb{H}$ is tall. Hence

$$\rho_n = 0, \ \lambda_n = \sigma_\eta^2, \ n = N + PL + 1, \cdots, ML. \quad (22)$$

At this case, $\lambda_1 = \rho_1 + \sigma_\eta^2 > \lambda_{ML} = \sigma_\eta^2$, and furthermore the minimum eigenvalue actually gives an estimation of the noise power. This property has been successfully used in system identification [23, 25] and direction of arrival (DOA) estimation (for example, see [21], page 656).

In practice, the number of source signals ($P$) and the channel orders usually are unknown, and therefore it is difficult to choose $L$ such that $L > N/(M - P)$. Moreover, to reduce complexity, we may only choose a small smoothing factor $L$ (may not satisfy $L > N/(M - P)$). At this case, if there is signal, it is possible that $\rho_{ML} \neq 0$. However, as explained above, it is almost sure that $\rho_1 > \rho_{ML}$ and therefore, $\lambda_1/\lambda_{ML} > 1$. Hence, we can almost always detect the signal existence by checking the ratio $\lambda_1/\lambda_{ML}$.

Let $\Delta$ be the average of all the eigenvalues of $\mathbf{R}_x$. For the same reason shown above, when there is no signal, $\Delta/\lambda_{ML} = 1$, and when there is signal, $\Delta/\lambda_{ML} > 1$. Hence, we can also detect if signal exists by checking the ratio $\Delta/\lambda_{ML}$.

The average eigenvalue $\Delta$ is almost the same as the signal energy (see the proof in the appendix B). Hence, we can use the ratio of the signal energy to the minimum eigenvalue for detection, which is the mathematical ground for the EME.

## 4 Performance Analysis and Detection Threshold

At finite number of samples, the sample covariance matrix $\mathbf{R}_x(N_s)$ may be well away from the statistical covariance matrices $\mathbf{R}_x$. The eigenvalue distribution of $\mathbf{R}_x(N_s)$ becomes very complicated [18, 19, 20, 21]. This makes the choice of the threshold very difficult. In this section, we will use some latest random matrix theories to set the threshold and obtain the probability of detection.

Let $P_d$ be the probability of detection, that is, at hypothesis $H_1$, the probability of the algorithm having detected the signal. Let $P_{fa}$ be the probability of false alarm, that is, at $H_0$, the probability of the algorithm having detected the signal. Since we have no information on the signal (actually we even do not know if there is signal or not), it is difficult to set the threshold based on the $P_d$. Hence, usually we choose the threshold based on the $P_{fa}$. The threshold is therefore not related to signal property and SNR.

## 4.1 Probability of false alarm and threshold

When there is no signal, $\mathbf{R}_x(N_s)$ turns to $\mathbf{R}_\eta(N_s)$, the sample covariance matrix of the noise defined as,

$$\mathbf{R}_\eta(N_s) = \frac{1}{N_s} \sum_{n=L-1}^{L-2+N_s} \hat{\boldsymbol{\eta}}(n)\hat{\boldsymbol{\eta}}^\dagger(n). \qquad (23)$$

$\mathbf{R}_\eta(N_s)$ is nearly a Wishart random matrix [18]. The study of the spectral (eigenvalue distributions) of a random matrix is a very hot topic in recent years in mathematics as well as communication and physics. The joint probability density function (PDF) of ordered eigenvalues of a Wishart random matrix has been known for many years [18]. However, since the expression of the PDF is very complicated, no closed form expression has been found for the marginal PDF of ordered eigenvalues. Recently, I. M. Johnstone and K. Johansson have found the distribution of the largest eigenvalue [19, 20] as described in the following theorem.

**Theorem 1**. Assume that the noise is real. Let $\mathbf{A}(N_s) = \frac{N_s}{\sigma_\eta^2}\mathbf{R}_\eta(N_s)$, $\mu = (\sqrt{N_s - 1} + \sqrt{ML})^2$ and $\nu = (\sqrt{N_s - 1} + \sqrt{ML})(\frac{1}{\sqrt{N_s-1}} + \frac{1}{\sqrt{ML}})^{1/3}$. Assume that $\lim_{N_s \to \infty} \frac{ML}{N_s} = y$ ($0 < y < 1$). Then $\frac{\lambda_{max}(\mathbf{A}(N_s)) - \mu}{\nu}$ converges (with probability one) to the Tracy-Widom distribution of order 1 ($W_1$) [26, 27].

Bai and Yin found the limit of the smallest eigenvalue [21] as described in the following theorem.

**Theorem 2**. Assume that $\lim_{N_s \to \infty} \frac{ML}{N_s} = y$ ($0 < y < 1$). Then $\lim_{N_s \to \infty} \lambda_{min} = \sigma_\eta^2(1 - \sqrt{y})^2$ (with probability one).

Based on the theorems, when $N_s$ is large, the largest and smallest eigenvalues of $\mathbf{R}_\eta(N_s)$ tend to deterministic values $\frac{\sigma_\eta^2}{N_s}(\sqrt{N_s} + \sqrt{ML})^2$ and $\frac{\sigma_\eta^2}{N_s}(\sqrt{N_s} - \sqrt{ML})^2$, respectively, that is, they are centered at the values, respectively, and have variances tend to zeros. Furthermore, Theorem 1 gives the distribution of the largest eigenvalue for large $N_s$.

The Tracy-Widom distributions were found by Tracy and Widom (1996) as the limiting law of the largest eigenvalue of certain random matrices [26, 27]. Let $F_1$ be the cumulative distribution function (CDF) (sometimes simply called distribution function) of the Tracy-Widom distribution of order 1. There is no closed form expression for the distribution function. The distribution function is defined as

$$F_1(t) = \exp\left(-\frac{1}{2}\int_t^\infty \left(q(u) + (u-t)q^2(u)\right) du\right), \qquad (24)$$

where $q(u)$ is the solution of the nonlinear Painlevé II differential equation

$$q''(u) = uq(u) + 2q^3(u). \qquad (25)$$

It is generally difficult to evaluate it. Fortunately, there have been tables for the functions [19] and Matlab codes to compute it [28]. Table 1 gives the values of $F_1$ at some points. It can also be used to compute the inverse $F_1^{-1}$ at certain points. For example, $F_1^{-1}(0.9) = 0.45$, $F_1^{-1}(0.95) = 0.98$.

Using the theories, we are ready to analyze the algorithms. The probability of false alarm of the MME detection is

$$\begin{aligned} P_{fa} &= P\left(\lambda_{max} > \gamma_1 \lambda_{min}\right) \\ &= P\left(\frac{\sigma_\eta^2}{N_s}\lambda_{max}(\mathbf{A}(N_s)) > \gamma_1 \lambda_{min}\right) \\ &\approx P\left(\lambda_{max}(\mathbf{A}(N_s)) > \gamma_1(\sqrt{N_s} - \sqrt{ML})^2\right) \\ &= P\left(\frac{\lambda_{max}(\mathbf{A}(N_s)) - \mu}{\nu} > \frac{\gamma_1(\sqrt{N_s} - \sqrt{ML})^2 - \mu}{\nu}\right) \\ &= 1 - F_1\left(\frac{\gamma_1(\sqrt{N_s} - \sqrt{ML})^2 - \mu}{\nu}\right). \end{aligned} \qquad (26)$$

This leads to

$$F_1\left(\frac{\gamma_1(\sqrt{N_s} - \sqrt{ML})^2 - \mu}{\nu}\right) = 1 - P_{fa}, \qquad (27)$$

or, equivalently,

$$\frac{\gamma_1(\sqrt{N_s} - \sqrt{ML})^2 - \mu}{\nu} = F_1^{-1}(1 - P_{fa}). \qquad (28)$$

From the definitions of $\mu$ and $\nu$, we finally obtain the threshold

$$\begin{aligned} \gamma_1 &= \frac{(\sqrt{N_s} + \sqrt{ML})^2}{(\sqrt{N_s} - \sqrt{ML})^2} \\ &\cdot \left(1 + \frac{(\sqrt{N_s} + \sqrt{ML})^{-2/3}}{(N_s ML)^{1/6}} F_1^{-1}(1 - P_{fa})\right). \end{aligned} \qquad (29)$$

**Please note that, unlike energy detection, here the threshold is not related to noise power. The threshold can be pre-computed based only on $N_s$, $L$ and $P_{fa}$, irrespective of signal and noise**.

Now we analyze the EME method. When there is no signal, it can be verified that the average energy defined in (7) satisfies

$$\mathrm{E}(T(N_s)) = \sigma_\eta^2, \ \mathrm{Var}(T(N_s)) = \frac{2\sigma_\eta^4}{MN_s}. \qquad (30)$$

$T(N_s)$ is the average of $MN_s$ statistically independent and identically distributed random variables. Since $N_s$ is large, the central limit theorem tells us that $T(N_s)$ can be approximated by the Gaussian distribution with mean $\sigma_\eta^2$ and

| $t$ | -3.90 | -3.18 | -2.78 | -1.91 | -1.27 | -0.59 | 0.45 | 0.98 | 2.02 |
|---|---|---|---|---|---|---|---|---|---|
| $F_1(t)$ | 0.01 | 0.05 | 0.10 | 0.30 | 0.50 | 0.70 | 0.90 | 0.95 | 0.99 |

Table 1: Numerical table for the Tracy-Widom distribution of order 1

variance $\frac{2\sigma_\eta^4}{MN_s}$. Hence the probability of false alarm is

$$P_{fa} = P(T(N_s) > \gamma_2 \lambda_{min})$$
$$\approx P\left(T(N_s) > \gamma_2 \frac{\sigma_\eta^2}{N_s}(\sqrt{N_s} - \sqrt{ML})^2\right)$$
$$= P\left(\frac{T(N_s) - \sigma_\eta^2}{\sqrt{\frac{2}{MN_s}}\sigma_\eta^2} > \frac{\gamma_2\sqrt{M}(\sqrt{N_s} - \sqrt{ML})^2 - \sqrt{M}N_s}{\sqrt{2N_s}}\right)$$
$$\approx Q\left(\frac{\gamma_2\sqrt{M}(\sqrt{N_s} - \sqrt{ML})^2 - \sqrt{M}N_s}{\sqrt{2N_s}}\right) \quad (31)$$

where

$$Q(t) = \frac{1}{\sqrt{2\pi}} \int_t^{+\infty} e^{-u^2/2} du. \quad (32)$$

Hence, we should choose the threshold such that

$$\frac{\gamma_2\sqrt{M}(\sqrt{N_s} - \sqrt{ML})^2 - \sqrt{M}N_s}{\sqrt{2N_s}} = Q^{-1}(P_{fa}). \quad (33)$$

That is,

$$\gamma_2 = \frac{Q^{-1}(P_{fa})\sqrt{2N_s} + \sqrt{M}N_s}{\sqrt{M}(\sqrt{N_s} - \sqrt{ML})^2}$$
$$= \left(\sqrt{\frac{2}{MN_s}}Q^{-1}(P_{fa}) + 1\right)\frac{N_s}{(\sqrt{N_s} - \sqrt{ML})^2}. \quad (34)$$

**Similar to MME, here the threshold is not related to noise power. The threshold can be pre-computed based only on $N_s$, $L$ and $P_{fa}$, irrespective of signal and noise.**

### 4.2 Probability of detection

When there is a signal, the sample covariance matrix $\mathbf{R}_x(N_s)$ is no longer a Wishart matrix. Up to now, the distributions of its eigenvalues are unknown. Hence, it is very difficult (mathematically intractable) to obtain a precisely closed form formula for the $P_d$. In this subsection, we try to approximate it and devise some empirical formulae.

Since $N_s$ is usually very large, we have the approximation

$$\mathbf{R}_x(N_s) \approx \mathbb{H}\mathbf{R}_s\mathbb{H}^\dagger + \mathbf{R}_\eta(N_s). \quad (35)$$

Note that $\mathbf{R}_\eta(N_s)$ approximates to $\sigma_\eta^2 \mathbf{I}_{ML}$. Hence, we have

$$\lambda_{max}(\mathbf{R}_x(N_s)) \approx \rho_1 + \lambda_{max}(\mathbf{R}_\eta(N_s)), \quad (36)$$
$$\lambda_{min}(\mathbf{R}_x(N_s)) \approx \rho_{ML} + \sigma_\eta^2. \quad (37)$$

For the MME method, the $P_d$ is

$$P_d = P(\lambda_{max}(\mathbf{R}_x(N_s)) > \gamma_1 \lambda_{min}(\mathbf{R}_x(N_s)))$$
$$\approx P(\lambda_{max}(\mathbf{R}_\eta(N_s)) > \gamma_1(\rho_{ML} + \sigma_\eta^2) - \rho_1)$$
$$= 1 - F_1\left(\frac{\gamma_1 N_s + N_s(\gamma_1 \rho_{ML} - \rho_1)/\sigma_\eta^2 - \mu}{\nu}\right) (38)$$

From the formula, the $P_d$ is related to the number of samples $N_s$, and the maximum and minimum eigenvalues of the signal covariance matrix (including channel effect).

Both the $P_d$ and threshold $\gamma_1$ in (29) are related to $L$ and $N_s$. For fixed $N_s$ and $P_{fa}$, the optimal $L$ is the one which maximizes the $P_d$. Based on (29) and (38), we can find that optimal $L$. However, the optimal $L$ does not have high practical value because it is related to signal property which is usually unknown at the receiver.

As proved in Appendix B,

$$T(N_s) = \frac{\text{Tr}(\mathbf{R}_x(N_s))}{ML}$$
$$\approx \frac{\text{Tr}(\mathbb{H}\mathbf{R}_s\mathbb{H}^\dagger)}{ML} + \frac{\text{Tr}(\mathbf{R}_\eta(N_s))}{ML}, \quad (39)$$

where $\text{Tr}(\cdot)$ means the trace of a matrix. As discussed in the last subsection, the minimum eigenvalue of $\mathbf{R}_\eta(N_s)$ is approximately $\frac{\sigma_\eta^2}{N_s}(\sqrt{N_s} - \sqrt{ML})^2$. Hence, equation (37) is an over-estimation for the minimum eigenvalue of $\mathbf{R}_x(N_s)$. On the other hand, $\rho_{ML} + \frac{\sigma_\eta^2}{N_s}(\sqrt{N_s} - \sqrt{ML})^2$ is obviously an under-estimation. Therefore, we choose an estimation between the two as

$$\lambda_{min}(\mathbf{R}_x(N_s)) \approx \rho_{ML} + \frac{\sigma_\eta^2}{\sqrt{N_s}}(\sqrt{N_s} - \sqrt{ML}). \quad (40)$$

Based on (39) and (40), we obtain an approximation for the $P_d$ of EME as

$$P_d = P(T(N_s) > \gamma_2 \lambda_{min}(\mathbf{R}_x(N_s))) \quad (41)$$
$$\approx P\left(\frac{\text{Tr}(\mathbf{R}_\eta(N_s))}{ML}\right.$$
$$\left. > \gamma_2\left(\rho_{ML} + \frac{\sigma_\eta^2}{\sqrt{N_s}}(\sqrt{N_s} - \sqrt{ML})\right) - \frac{\text{Tr}(\mathbb{H}\mathbf{R}_s\mathbb{H}^\dagger)}{ML}\right)$$
$$= Q\left(\frac{\gamma_2\left(\rho_{ML} + \frac{\sigma_\eta^2}{\sqrt{N_s}}(\sqrt{N_s} - \sqrt{ML})\right) - \frac{\text{Tr}(\mathbb{H}\mathbf{R}_s\mathbb{H}^\dagger)}{ML} - \sigma_\eta^2}{\sqrt{\frac{2}{MN_s}}\sigma_\eta^2}\right).$$

From the formula, the $P_d$ is related to the number of samples $N_s$, and the average and minimum eigenvalues of the signal covariance matrix (including channel effect).

Similarly, for fixed $N_s$ and $P_{fa}$, we can find the optimal $L$ based on (34) and (41).

## 4.3 Computational complexity

The major complexity of MME and EME comes from two parts: computation of the covariance matrix (equation (21)) and the eigenvalue decomposition of the covariance matrix. For the first part, noticing that the covariance matrix is a block Toeplitz matrix and Hermitian, we only needs to evaluate its first block row. Hence $M^2LN_s$ multiplications and $M^2L(N_s-1)$ additions are needed. For the second part, generally $O((ML)^3)$ multiplications and additions are sufficient. The total complexity (multiplications and additions, respectively) are therefore as follows:

$$M^2LN_s + O(M^3L^3). \qquad (42)$$

Since $N_s$ is usually much larger than $L$, the first part is dominate.

The energy detection needs $MN_s$ multiplications and $M(N_s-1)$ additions. Hence, the complexity of the proposed methods is about $ML$ times that of the energy detection.

## 5 Simulations and Discussions

In the following, we will give some simulation results using the randomly generated signals, wireless microphone signals [29] and the captured DTV signals [30].

### 5.1 Simulations

We define the SNR as the ratio of the average received signal power to the average noise power

$$\text{SNR} \overset{\text{def}}{=} \frac{E(||\mathbf{x}(n) - \boldsymbol{\eta}(n)||^2)}{E(||\boldsymbol{\eta}(n)||^2)}. \qquad (43)$$

We require the probability of false alarm $P_{fa} \leqslant 0.1$. Then the threshold is found based on the formulae in Section IV. For comparison, we also simulate the energy detection with or without noise uncertainty for the same system. The threshold for the energy detection is given in [6]. At noise uncertainty case, the threshold is always set based on the assumed/estimated noise power, while the real noise power is varying in each Monte Carlo realization to a certain degree as specified by the noise uncertainty factor defined in Section II.

(1) **Multiple-receiver signal detection**. We consider a 2-input 4-receiver system ($M = 4$, $P = 2$) as defined by (6). The channel orders are $N_1 = N_2 = 9$ (10 taps). The channel taps are random numbers with Gaussian distribution. All the results are averaged over 1000 Monte Carlo realizations (for each realization, random channel, random noise and random BPSK inputs are generated).

For fixed $L = 8$ and $N_s = 100000$, the $P_d$ for the MME and energy detection (with or without noise uncertainty) are shown in Figure 1, where and in the following "EG-x dB" means the energy detection with x-dB noise uncertainty. If the noise variance is exactly known ($B = 0$), the energy detection is very good (note that it is optimal for iid signal). The proposed methods are slightly worse than the energy detection with ideal noise power. However, as discussed in [9, 6, 31], noise uncertainty is always present. As shown in the figure, if there is 0.5 to 2 dB noise uncertainty, the detection probability of the energy detection is much worse than that of the proposed methods. From the figure, we see that the theoretical formulae in Section IV.B for the $P_d$ (the curves with mark "MME-theo" and "EME-theo") are somewhat conservative.

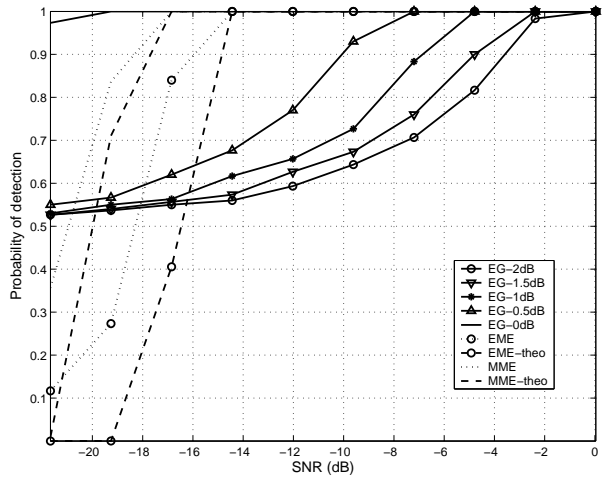

Figure 1: Probability of detection: $M = 4$, $P = 2$, $L = 8$.

The $P_{fa}$ is shown in Table 2 (second row) (note that $P_{fa}$ is not related to the SNR because there is no signal). The $P_{fa}$ for the proposed methods and the energy detection without noise uncertainty almost meet the requirement ($P_{fa} \leqslant 0.1$), but the $P_{fa}$ for the energy detection with noise uncertainty far exceeds the limit. This means that the energy detection is very unreliable in practical situations with noise uncertainty.

To test the impact of the number of samples, we fix the SNR at -20dB and vary the number of samples from 40000 to 180000. Figure 2 and Figure 3 show the $P_d$ and $P_{fa}$, respectively. It is seen that the $P_d$ of the proposed algorithms and the energy detection without noise uncertainty increases with the number of samples, while that of the energy detection with noise uncertainty almost does not change (this phenomenon is also verified in [10, 17]). This means that the noise uncertainty problem cannot be solved by increasing the number of samples. For the $P_{fa}$, all the algorithms do not change much with varying number of samples.

To test the impact of the smoothing factor, we fix the SNR at -20dB, $N_s = 130000$ and vary the smoothing factor $L$ from 4 to 14. Figure 4 shows the results for both $P_d$ and $P_{fa}$. It is seen that both $P_d$ and $P_{fa}$ of the pro-posed algorithms slightly increase with $L$, but will reach a ceiling at some $L$. Even if $L \leqslant N/(M-P) = 9$, the methods still works well (much better than the energy detection with noise uncertainty). Noting that smaller $L$ means lower complexity, in practice, we can choose a relatively small $L$. However, it is very difficult to choose the best $L$ because it is related to signal property (unknown). Note that the $P_d$ and $P_{fa}$ for the energy detection do not change with $L$.

| method | EG-2 dB | EG-1.5 dB | EG-1 dB | EG-0.5 dB | EG-0dB | EME | MME |
|---|---|---|---|---|---|---|---|
| $P_{fa}$ ($M=4$, $P=2$, $L=8$, $N_s=10^5$) | 0.499 | 0.499 | 0.498 | 0.495 | 0.104 | 0.065 | 0.103 |
| $P_{fa}$ ($M=P=1$, $L=10$, $N_s=5\times 10^4$) | 0.497 | 0.496 | 0.488 | 0.470 | 0.107 | 0.019 | 0.074 |
| $P_{fa}$ ($M=2$, $P=1$, $L=8$, $N_s=5\times 10^4$) | 0.499 | 0.499 | 0.497 | 0.486 | 0.097 | 0.028 | 0.072 |

Table 2: Probability of false alarm at different parameters

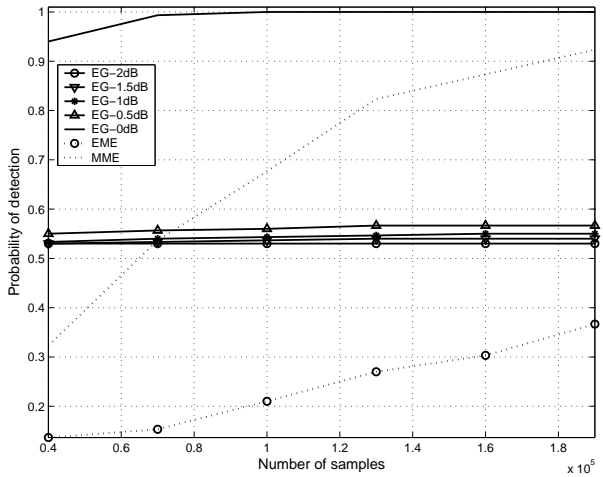

Figure 2: Probability of detection: $M=4$, $P=2$, $L=8$, SNR=$-20$ dB.

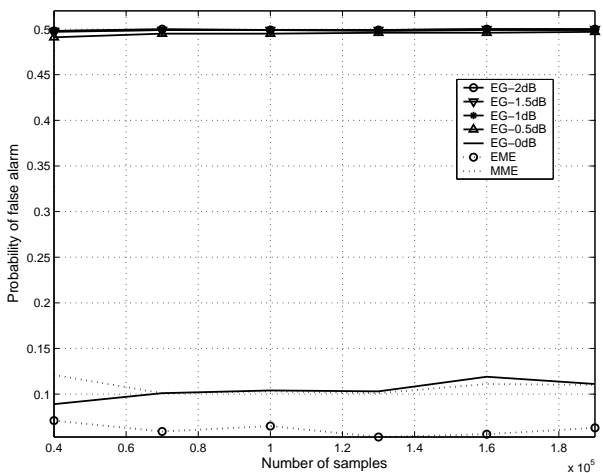

Figure 3: Probability of false alarm: $M=4$, $P=2$, $L=8$.

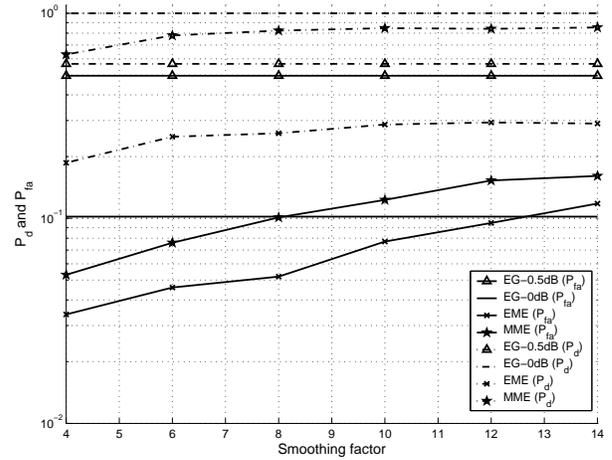

Figure 4: Impact of smoothing factor: $M=4$, $P=2$, SNR=$-20$ dB, $N_s=130000$.

(2) **Wireless microphone signal detection**. FM modulated analog wireless microphone is widely used in USA and elsewhere. It operates on TV bands and typically occupies about 200KHz (or less) bandwidth [29]. The detection of the signal is one of the major challenge in 802.22 WRAN [5]. In this simulation, wireless microphone soft speaker signal [29] at central frequency $f_c = 200$ MHz is used. The sampling rate at the receiver is 6 MHz (the same as the TV bandwidth in USA). The smoothing factor is chosen as $L = 10$. Simulation results are shown in Figure 5 and Table 2 (third row for $P_{fa}$). From the figure and the table, we see that all the claims above are also valid here. Furthermore, here the MME is even better than the ideal energy detection. The reason is that here the signal samples are highly

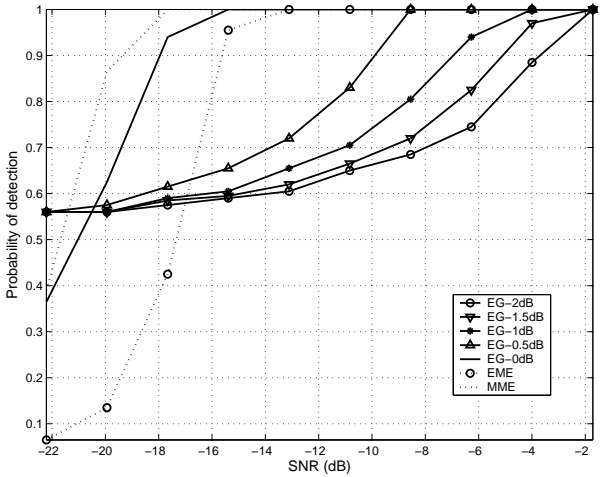

Figure 5: Probability of detection for wireless microphone signal.

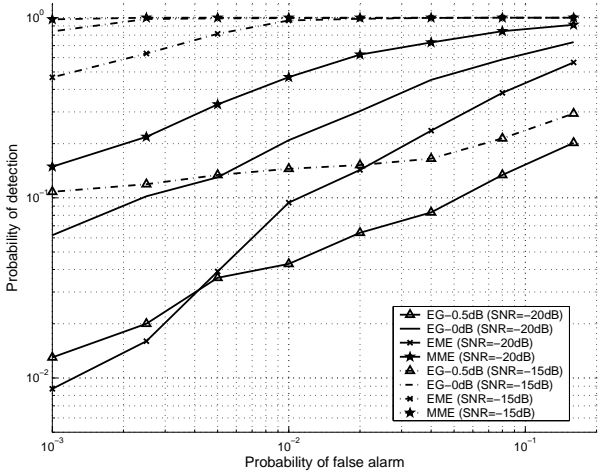

Figure 6: ROC curve for wireless microphone signal: $N_s = 50000$.

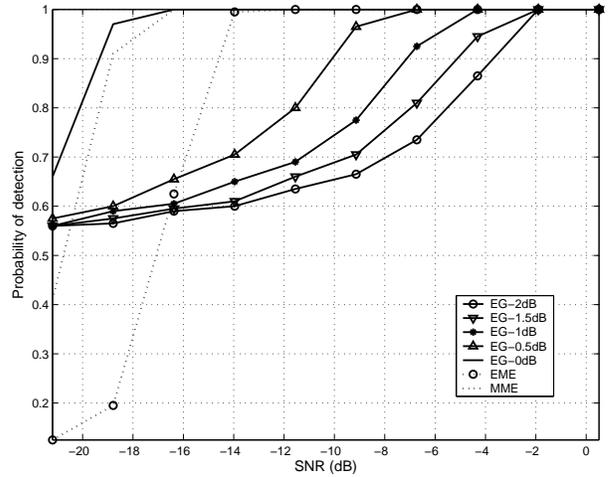

Figure 7: Probability of detection for DTV signal WAS-003/27/01.

correlated and therefore energy detection is not optimal.

The Receiver Operating Characteristics (ROC) curve is shown in Figure 6, where the sample size is $N_s = 50000$. Note that we need slightly adjusting the thresholds to keep all the methods having the same $P_{fa}$ values (especially for the energy detection with noise uncertainty, the threshold based on the predicted noise power and theoretical formula is very inaccurate to obtain the target $P_{fa}$ as shown in Table 2). It shows that MME is the best among all the methods. The EME is worse than the ideal energy detection but better than the energy detection with noise uncertainty 0.5 dB.

(3) **Captured DTV signal detection**. Here we test the algorithms based on the captured ATSC DTV signals [30]. The real DTV signals (field measurements) are collected at Washington D.C. and New York, USA, respectively. The sampling rate of the vestigial sideband (VSB) DTV signal is 10.762 MHz [32]. The sampling rate at the receiver is two times that rate (oversampling factor is 2). The multipath channel and the SNR of the received signal are unknown. In order to use the signals for simulating the algorithms at very low SNR, we need to add white noises to obtain various SNR levels [31]. In the simulations, the smoothing factor is chosen as $L = 8$. The number of samples used for each test is $2N_s = 100000$ (corresponding to 4.65 ms sampling time). The results are averaged over 1000 tests (for different tests, different data samples and noise samples are used). Figure 7 gives the $P_d$ based on the DTV signal file WAS-003/27/01 (at Washington D.C., the receiver is outside and 48.41 miles from the DTV station) [30]. Figure 8 gives the results based on the DTV signal file NYC/205/44/01 (at New York, the receiver is indoor and 2 miles from the DTV station) [30]. Note that each DTV signal file contains data samples in 25 seconds. The $P_{fa}$ are shown in Table 2 (fourth row). The simulation results here are similar to those for the randomly generated signals.

In summary, all the simulations show that the proposed methods work well without using the information of signal, channel and noise power. The MME is always better than the EME (yet theoretical proof has not been found). The energy detection are not reliable (low probability of detection and high probability of false alarm) when there is noise uncertainty.

## 5.2 Discussions

Theoretic analysis of the proposed methods highly relies on the random matrix theory, which is currently one of the hot topic in mathematics as well as in physics and communication. We hope advancement on the random matrix theory can solve the following open problems.

(1) Accurate and analytic expression for the $P_d$ at given

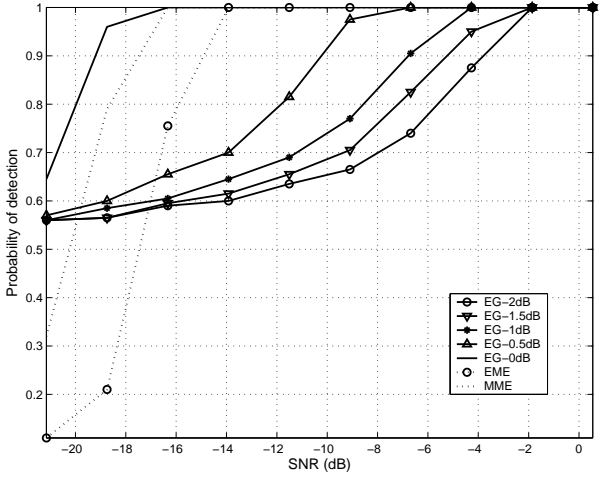

Figure 8: Probability of detection for DTV signal NYC/205/44/01.

threshold. This requires the eigenvalue distribution of matrix $\mathbf{R}_x(N_s)$ when both signal and noise are present. At this case, $\mathbf{R}_x(N_s)$ is no longer a Wishart random matrix.

(2) When there is no signal, the exact solution of $P(\lambda_{max}/\lambda_{min} > \gamma)$. This is the $P_{fa}$. That is, we need to find the distribution of $\lambda_{max}/\lambda_{min}$. As we know, this is still an unsolved problem. In this paper, we have approximated this probability through replacing $\lambda_{min}$ by a deterministic number.

(3) Strictly speaking, the sample covariance matrix of the noise $\mathbf{R}_\eta(N_s)$ is not a Wishart random matrix, because the $\hat{\boldsymbol{\eta}}(n)$ for different $n$ are correlated. Although the correlations are weak, the eigenvalue distribution may be affected. Is it possible to obtain a more accurate eigenvalue distribution by using this fact?

# 6 Conclusions

Methods based on the eigenvalues of the sample covariance matrix of the received signal have been proposed. Latest random matrix theories have been used to set the thresholds and obtain the probability of detection. The methods can be used for various signal detection applications without knowledge of signal, channel and noise power. Simulations based on randomly generated signals, wireless microphone signals and captured DTV signals have been done to verify the methods.

# Appendix A

At the receiving end, usually the received signal is filtered by a narrowband filter. Therefore, the noise embedded in the received signal is also filtered. Let $\eta(n)$ be the noise samples before the filter, which are assumed to be independent and identically distributed (i.i.d). Let $f(k)$, $k = 0, 1, \cdots, K$, be the filter. After filtering, the noise samples turns to

$$\tilde{\eta}(n) = \sum_{k=0}^{K} f(k)\eta(n-k), \ n = 0, 1, \cdots . \quad (44)$$

Consider $L$ consecutive outputs and define

$$\tilde{\boldsymbol{\eta}}(n) = [\tilde{\eta}(n), \cdots, \tilde{\eta}(n - L + 1)]^T. \quad (45)$$

The statistical covariance matrix of the filtered noise becomes

$$\tilde{\mathbf{R}}_\eta = \mathrm{E}(\tilde{\boldsymbol{\eta}}(n)\tilde{\boldsymbol{\eta}}(n)^\dagger) = \sigma_\eta^2 \mathbf{H}\mathbf{H}^\dagger, \quad (46)$$

where $\mathbf{H}$ is a $L \times (L+K)$ matrix defined as

$$\mathbf{H} = \begin{bmatrix} f(0) & f(1) & \cdots & f(K) & 0 & \cdots & 0 \\ 0 & f(0) & \cdots & f(K-1) & f(K) & \cdots & 0 \\ & & \ddots & & & \ddots & \\ 0 & 0 & \cdots & f(0) & f(1) & \cdots & f(K) \end{bmatrix}. \quad (47)$$

Let $\mathbf{G} = \mathbf{H}\mathbf{H}^\dagger$. If analog filter or both analog and digital filters are used, the matrix $\mathbf{G}$ should be defined based on those filter properties. Note that $\mathbf{G}$ is a positive definite Hermitian matrix. It can be decomposed to $\mathbf{G} = \mathbf{Q}^2$, where $\mathbf{Q}$ is also a positive definite Hermitian matrix. Hence, we can transform the statistical covariance matrix into

$$\mathbf{Q}^{-1}\tilde{\mathbf{R}}_\eta \mathbf{Q}^{-1} = \sigma_\eta^2 \mathbf{I}_L. \quad (48)$$

Note that $\mathbf{Q}$ is only related to the filter. This means that we can always transform the statistical covariance matrix $\mathbf{R}_x$ in (17) (by using a matrix obtained from the filter) such that equation (20) holds when the noise has been passed through a narrowband filter. Furthermore, since $\mathbf{Q}$ is not related to signal and noise, we can pre-compute its inverse $\mathbf{Q}^{-1}$ and store it for later usage.

# Appendix B

It is known that the summation of the eigenvalues of a matrix is the trace of the matrix. Let $\Delta(N_s)$ be the average of the eigenvalues of $\mathbf{R}_x(N_s)$. Then

$$\begin{aligned} \Delta(N_s) &= \frac{1}{ML}\mathrm{Tr}(\mathbf{R}_x(N_s)) \\ &= \frac{1}{MLN_s} \sum_{n=L-1}^{L-2+N_s} \hat{\mathbf{x}}^\dagger(n)\hat{\mathbf{x}}(n). \end{aligned} \quad (49)$$

After some mathematical manipulations, we obtain

$$\Delta(N_s) = \frac{1}{MLN_s} \sum_{i=1}^{M} \sum_{m=0}^{L-2+N_s} \delta(m)|x_i(m)|^2, \quad (50)$$

where

$$\delta(m) = \begin{cases} m+1, & 0 \leq m \leq L-2 \\ L, & L-1 \leq m \leq N_s-1 \\ N_s+L-m-1, & N_s \leq m \leq N_s+L-2 \end{cases} \quad (51)$$

Since $N_s$ is usually much larger than $L$, we have

$$\Delta(N_s) \approx \frac{1}{MN_s} \sum_{i=1}^{M} \sum_{m=0}^{N_s-1} |x_i(m)|^2 = T(N_s). \quad (52)$$